\documentclass[prl,twocolumn,showpacs,superscriptaddress]{revtex4}

\usepackage[dvips]{graphicx}
\usepackage{color}
\usepackage{latexsym}
\usepackage{amsmath}
\usepackage{amsthm}
\usepackage{amsfonts}
\usepackage{amssymb}

\newcommand{\rem}[1]{}

\newcommand{\beq}{\begin{equation}}
\newcommand{\eeq}{\end{equation}}
\newcommand{\beqa}{\begin{eqnarray}}
\newcommand{\eeqa}{\end{eqnarray}}

\begin{document}
\title{
Resonant magneto-conductance through a vibrating nanotube
}

\author{G. Rastelli}
\affiliation{
Laboratoire de Physique et Mod{\'e}lisation des Milieux Condens{\'e}s\\
Universit{\'e} Joseph Fourier \& CNRS, F-38042 Grenoble, France}

\author{M. Houzet}
\affiliation{CEA, INAC, SPSMS, F-38054 Grenoble, France}

\author{F. Pistolesi}
\affiliation{
Laboratoire de Physique et Mod{\'e}lisation des Milieux Condens{\'e}s\\
Universit{\'e} Joseph Fourier \& CNRS, F-38042 Grenoble, France}
\affiliation{CPMOH, Universit\'e de Bordeaux I \& CNRS, F-33405 Talence, France}

\date{\today}

\begin{abstract}
We address the electronic resonant transport in presence of a transverse magnetic field
through the single level of a suspended carbon nanotube acting as a quantum oscillator.
We predict a negative magneto-conductance with a
magnetic-field induced narrowing of the resonance line
and a reduction of the conductance peak when the nanotube
is asymmetrically contacted to the leads.
At finite bias voltage we study the threshold for phonon-assisted transport.
\end{abstract}

\pacs{
73.63.-b, 
71.38.-k, 
73.63.Kv, 
85.85.+j. 
}
\maketitle

A distinctive non-classical feature of any quantum state is
the possibility that part of the system is spatially delocalized.
Probing delocalization in macroscopic systems
is important to validate quantum mechanics on that scale.
For large molecules, this has been done by the observation of interference fringes with
diffraction experiments~\cite{arndt:1999}.
Is it possible to observe quantum delocalization for a mechanical oscillator?
Recently, it has been predicted that the magneto-conductance can be used as a detector
of the quantum delocalization of a suspended carbon nanotube due to the Aharonov-Bohm effect
on the electrons crossing the device~\cite{shekhter:2006}.
This spectacular prediction was done in the tunnel regime.
Though simple and transparent from the technical point of view,
this regime is not optimal for the experimental
observation of the effect for two reasons:
({\em i}) the current is very low,
({\em ii}) electrons can interfere only once since a single crossing through
the device contributes to the current.

In this article, we consider the resonant transport
through a single electron state of the nanotube.
At resonance, the electron channel is fully open for a static nanotube.
One could expect that the magneto-conductance signal for the suspended nanotube will be greatly enhanced:
The electrons bounce many times inside the structure before
leaving, therefore allowing multiple interference.
Thus, even if the phase acquired at each passage is small, the accumulated phase can be large.
By performing a calculation with Keldysh non-equilibrium Green's function technique
at lowest order in the electron-phonon coupling,
we find the following results:
The shape of the resonance as a function of the gate voltage is
modified by the magnetic field.
At resonance and for vanishing temperature, the linear conductance
depends on the magnetic field only if the coupling to the leads
is asymmetric, while the resonance width is reduced
by the magnetic field.
The current-voltage characteristics shows
a magnetic-field-dependent singularity at the threshold of
one-phonon absorption.
These prominent features constitute
a measurable signature of the quantum delocalization of the
nanotube.

We model the system with the following Hamiltonian:
\begin{equation} \label{eq:Hamiltonian}
    H
    =
    \sum_{\nu=l,r}\sum_{k}
    {\xi}^{\vphantom{\dagger}}_{k\nu} a^\dagger_{k\nu} {a}^{\vphantom{\dagger}}_{k\nu}
    + {\varepsilon}^{\vphantom{\dagger}}_d \, {a}^{\dagger}_d  {a}^{\vphantom{\dagger}}_d
    + \hbar \omega  \, {b}^{\dagger}  {b}^{\vphantom\dagger} +  H_T
    \, .
\end{equation}
Here,  $a^\dagger_{kl}$ ($a^\dagger_{kr}$)
is a creation operator for the electronic single particle
states in the left (right) lead.
The energy spectrum in each lead is $\xi_{k\nu}=\varepsilon_k-\mu_\nu$,
where the difference of the chemical potentials $\mu_r-\mu_l=eV$ is related
to the bias voltage $V$.
The leads are connected by a suspended nanotube placed
in a strong magnetic field perpendicular to the nanotube's oscillation plane.
We single out the fundamental bending mode with resonance frequency $\omega$
for which $b^\dag$ is the creation operator of quantum excitations.
The single relevant electronic level in the nanotube for which $a^\dagger_d$ is
a creation operator sits at energy $\varepsilon_d$ that can be controlled with an
external gate voltage.
For simplicity, we consider fully spin-polarized electrons (having in mind
the large magnetic field for which our theory applies) and neglect a possible
orbital degeneracy in the nanotube.
The last term in Eq.~(\ref{eq:Hamiltonian}) models electron transfer
from the leads to the nanotube in presence of the magnetic field~\cite{shekhter:2006}:
\begin{equation}\label{eq:HT}
    H_T =
    \sum_{\nu=l,r}\sum_k
    {t}^{\vphantom{\dagger}}_\nu e^{i{\alpha_\nu} x} a^{\dagger}_{k\nu} {a}^{\vphantom{\dagger}}_d
    +
    \mbox{h.c.}
    \, .
\end{equation}
Here, $x=b+b^\dagger$ is the displacement operator of the nanotube in units of
the amplitude $u_0$ for zero-point fluctuations, $t_l$ and $t_r$
are the tunneling matrix elements at the point contacts between
the nanotube and the leads.
The factor $\alpha\equiv\alpha_l=-\alpha_r= g B L u_0 / \Phi_0$ is the magnetic flux
(in units of the flux quantum $\Phi_0=h/e$)
through the area swept by the nanotube with length $L$
in the ground-state quantum fluctuation at magnetic field $B$
($g$ is a numerical factor of order one).
Typically, $\alpha$ is very small: for a single
wall nanotube of length $L = 1 \mu$m,
$(\omega/2\pi)= 500$ MHz, one finds
$u_0 \simeq 1$ pm.
With $B = 40$T, one obtains $\alpha \sim 0.1$.
The Hamiltonian (\ref{eq:Hamiltonian}) resembles the one proposed to
study polaronic transport through a vibrating molecule~\cite{glazman:1988,wingreen:1988}
(there $\alpha_l=\alpha_r$, $\varepsilon_d$ includes a polaronic shift $-\alpha^2\omega$,
and $x$ stands for dimensionless momentum),
but they lead to qualitative different behavior.
The Hamiltonian of Ref. \cite{imam:1994} describing Coulomb blockade at resonant tunneling
reduces to (\ref{eq:Hamiltonian}) in some limit.
We address a regime that was not considered there.

The current operator in the left lead is $\hat{I}\equiv (i e/\hbar) [H, N_l  ]$,
with $N_l=\sum_{k} {a}^{\dagger}_{kl}  {a}^{\vphantom{\dagger}}_{kl}$.
In the stationary regime, the dc current flowing through the device is
\begin{equation} \label{eq:current}
I=-\frac{ie}{\hbar}\sum_{k}
  \left(  {t}^{\vphantom{\dagger}}_l \langle e^{i{\alpha} x} a^{\dagger}_{kl} {a}^{\vphantom{\dagger}}_d \rangle
    - \mbox{c.c.}
    \right)\,,
\end{equation}
where the brackets denote a quantum-statistical average.
In order to evaluate the current, we use the Keldysh theory for
non-equilibrium systems~\cite{rammer-smith:1986}.
We define the retarded, advanced,
and Keldysh electronic Green's functions:
$G^{R/A}_{n,n'}(t)=\mp i \theta(\pm t)\langle\{a_n(t),a_{n'}^\dagger(0)\}\rangle$,
$G^{K}_{n,n'}(t)=-i \langle[a_n(t),a_{n'}^\dagger(0)]\rangle$
(with $n,n'=kl,k'r,d$) and we
build a triangular 2X2 matrix $\hat{G}$ out of them, with elements
$\hat{G}_{11}=G^R$, $\hat{G}_{12}=G^K$, $\hat{G}_{22}=G^A$,
and $\hat{G}_{21}=0$.
We define similarly a Green's function $\hat{G}_{xd,kl}$ related to Eq.~(\ref{eq:current}),
such that, for instance,
$G_{xd,kl}^{R/A}(t)=\mp i \theta(\pm t)\langle\{a_d(t)e^{i\alpha
x(t)},a_{kl}^\dagger(0)\}\rangle$.
The relation
$\hat{G}_{kl,k'l}(\varepsilon)=\hat{g}_{kl}(\varepsilon)\delta_{k,k'}+\hat{g}_{kl}(\varepsilon)
t_l\hat{G}_{xd,k'l}(\varepsilon)
$
holds in Fourier space, where $\hat{g}_{kl}$  is the Green's function in the
uncoupled left lead (at $H_T=0$).
We introduce $\hat{g}_{\nu}=\sum_{k}\hat{g}_{k\nu}$ and
$\hat{G}_{\nu,\nu'}=\sum_{k,k'}\hat{G}_{k\nu,k'\nu'}$.
In the wide-band limit: $g_\nu^{R/A}=\mp i\pi \rho_{\nu}$,
$g_\nu^K=2[1-2n_\nu]g_\nu^{R}$,
where $\rho_\nu$ are the densities of states in the leads and $n_\nu(\varepsilon)=n_F(\varepsilon-\mu_\nu)$,
with $n_F$ the Fermi distribution function at temperature $T$.
Then, Eq.~(\ref{eq:current}) can be rewritten:
\begin{equation} \label{eq:current2}
I=-\frac{e}{h}\mbox{Re}\int d\varepsilon
\left[\hat{g}_l^{-1}(\varepsilon)\hat{G}_{l,\, l}(\varepsilon)
\right]_{12}.
\end{equation}

In the absence of electron-phonon coupling (at $\alpha=0$), the Green's function on the dot is known:
$G_{d,\,d}^{(0)R/A}(\varepsilon)=(\varepsilon-\varepsilon_d\pm i \Gamma)^{-1}$, while
$G_{d,\,d}^{(0)K}(\varepsilon)=-2i[(\varepsilon-\varepsilon_d)^2+ \Gamma^2]^{-1}\sum_{\nu=l,r}\Gamma_\nu[1-2n_\nu(\varepsilon)]$.
Here, $\Gamma=\Gamma_l+\Gamma_r$ and $\Gamma_{l/r}=\pi \rho_{l/r} |t_{l/r}|^2$ give
the broadening of the resonant level due to its hybridization with the leads.
Then, one gets:
$\hat{G}_{l,\,l}^{(0)}=\hat{g}_{l}+\hat{g}_{l}t_l\hat{G}_{d,d}^{(0)}t_l^*\hat{g}_{l}$.
Inserting this into Eq.~(\ref{eq:current2}), one retrieves the result
\begin{equation} \label{eq:I0}
I^{(0)}=\frac{e}{h}\int d\varepsilon
[n_l(\varepsilon)-n_r(\varepsilon)]{\cal T}(\varepsilon),
\end{equation}
with the elastic Breit-Wigner transmission coefficient through the non-interacting resonant level
\begin{equation} \label{eq:transmission}
        {\cal T}(\varepsilon)
        =
        \frac{4 \Gamma_l\Gamma_r}{(\varepsilon-\varepsilon_d)^2+\Gamma^2} \, .
\end{equation}
In particular, the linear conductance at resonance and $T=0$, $G_\mathrm{max}=(e^2/h)4\Gamma_l\Gamma_r/\Gamma^2$, reaches the conductance quantum for symmetric contacts ($\Gamma_l=\Gamma_r$).

\begin{figure}[htbp]
\includegraphics[scale=0.18,angle=0.]{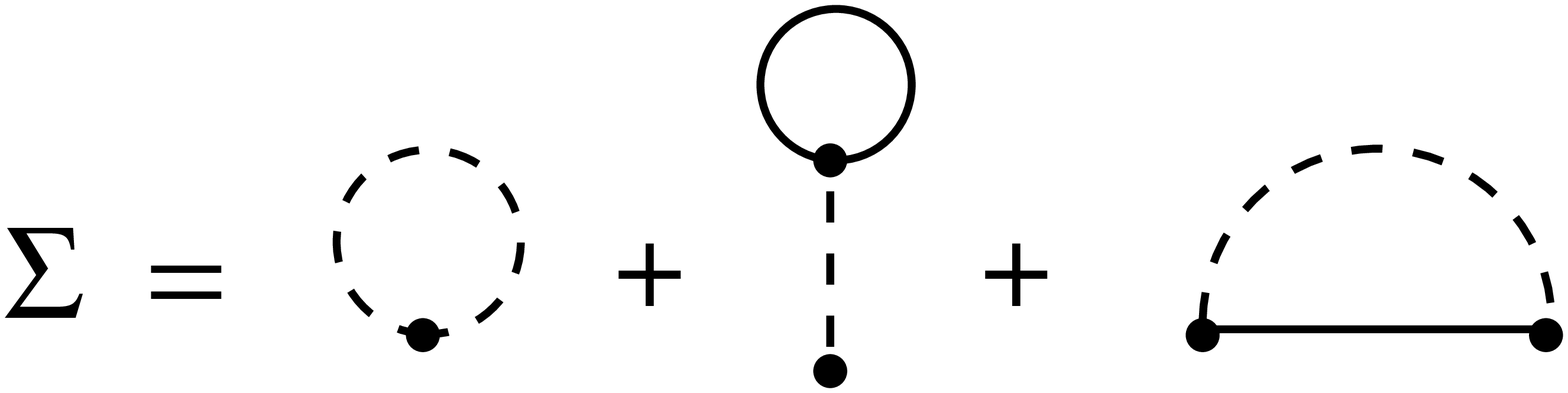}
\caption{
Diagrams for the self-energy. Dots denote vertices at tunneling from the single level
to the leads, full and dashed lines stand for electron and phonon Green's functions, respectively.}
\label{fig1}
\end{figure}

We now consider the coupling with phonons perturbatively.
To lowest order in $\alpha$, the Green's functions read:
\begin{equation}\label{eq:green2}
\hat{G}_{l,\,l}(\varepsilon)=\hat{G}^{(0)}_{l,\,l}(\varepsilon)+\sum_{n,n'=l,r,d}\hat{G}^{(0)}_{l,n}(\varepsilon)\hat{\Sigma}^{(2)}_{n,n'}(\varepsilon)\hat{G}^{(0)}_{n',l}(\varepsilon).
\end{equation}
The self-energies $\hat{\Sigma}^{(2)}_{n,n'}$ are represented schematically
in Fig.~\ref{fig1} by one-loop diagrams.
%
%
%
The first diagram arises from the terms proportional to $\alpha^2 x^2$ in the perturbative
expansion of Eq.~(\ref{eq:HT}) with respect to $\alpha$.
It leads to a renormalization of tunneling matrix elements:
$t_{l/r}\rightarrow t_{l/r}(1-\alpha^2\langle x^2 \rangle_0/2)$ , where
$\langle x^2 \rangle_0=\coth[\hbar\omega/(2k_B T)]$ for the unperturbed oscillator.
The second diagram accounts for the shift of the oscillator's position
due to the Lorentz force acting on it: $\langle x \rangle_\alpha=4 \alpha  I^{(0)}/(e\omega)$.
It does not contribute to the current to order $\alpha^2$.
The third diagram contains the non-trivial part of the electron-phonon interaction.
The sum of the three diagrams reads:
\begin{equation}
    \hat\Sigma_{n,n'}^{(2)}
    =
    {\tilde t}_{nn'}+\sum_{m, m'=l,r,d} A_{nn'}^{m m'} \hat \sigma_{m, m'},
\end{equation}
where
${\tilde t}_{l/r,d}= {\tilde t}_{d,l/r}^* = t_{l/r} (\pm i \alpha \langle x \rangle_\alpha-\alpha^2\langle x^2 \rangle_0/2  )$,
$A_{\nu \nu'}^{dd}={A_{dd}^{\nu \nu'}}^*=\alpha_\nu \alpha_{\nu'} t_\nu t_{\nu'}^*$,
$A_{\nu d}^{d \nu'}={A_{d\nu'}^{\nu d}}^*=-\alpha_\nu \alpha_{\nu'} t_\nu t_{\nu'}$
($\nu,\nu'=l,r$) and zero otherwise.
The components of $\hat\sigma_{n,n'}$ ($n,n'=l,r,d$) read
\begin{eqnarray}
\label{eq:reduct-self}
\sigma_{n,n'}^{R/A}(t)&=&
\frac{i}{2}
\left[G_{n,n'}^{(0)R/A}(t)D^{K}(t)+G_{n,n'}^{(0)K}(t)D^{R/A}(t)
\right] ,
\nonumber
\\
\sigma_{n,n'}^K(t)&=&
\frac{i}{2}
\left[G_{n,n'}^{(0)R}(t)D^{R}(t)
+G_{n,n'}^{(0)A}(t)D^{A}(t)
\right.
\nonumber \\
&&
\left.
+G_{n,n'}^{(0)K}(t)D^K(t)
\right].
\end{eqnarray}
Here, $\hat{D}$ is the Green's function for unperturbed phonons (at $\alpha=0$):
$D^{R/A}(\varepsilon)=2\hbar\omega/[(\varepsilon\pm i0^+)^2-(\hbar\omega)^2]$
and
$D^K(\varepsilon)=-2i\pi[\delta(\varepsilon-\hbar\omega)+\delta(\varepsilon+\hbar\omega)]\coth[\hbar\omega/(2 k_B T)]$.

Evaluation of the current up to $\alpha^2$ terms by inserting
eqs.~(\ref{eq:green2})-(\ref{eq:reduct-self}) into (\ref{eq:current2}) is now straightforward.
After lengthy calculations, we get $I = I^{(0)} + \alpha^2 I^{(2)}+\dots$, where
\begin{eqnarray} \label{eq:I2}
&&\lefteqn{I^{(2)} = \frac{G_\mathrm{max}}{e}
\left[
\int \!\! d\varepsilon
 \left[ n_l(\varepsilon) - n_r(\varepsilon) \right] {\cal T}^{a}(\varepsilon) \right.}
 \nonumber \\
&+&
 \sum_{\nu,\nu'=l,r} \left(
\int d\varepsilon
\left[ 1 - n_{\nu'}(\varepsilon-\hbar\omega) \right]  n_{\nu}(\varepsilon)  {\cal T}^{b}_{\nu,\nu'}(\varepsilon)
\right.
\nonumber \\
&+&
\left.\left.  \int \int   d\varepsilon  d\varepsilon'
 \left[ 1 - n_{\nu'}(\varepsilon') \right]  n_{\nu}(\varepsilon)
 {\cal T}^{c}_{\nu,\nu'}(\varepsilon,\varepsilon') \right)
\right].
\end{eqnarray}
The coefficients
\begin{widetext}
\begin{subequations}
\label{eq:T}
\begin{eqnarray}
\!\!\!\! {\cal T}^{a}(\varepsilon) &=& \!\!\! -
2 a\left(\varepsilon\right) \left[ a(\varepsilon+\hbar \omega) + a(\varepsilon)  \right]
- 4 \gamma^2 b(\varepsilon)  c(\varepsilon,\varepsilon+\hbar \omega)
-   \left[2 b(\varepsilon) - b(\varepsilon+\hbar\omega)  - b(\varepsilon-\hbar\omega)  \right] n_B
\label{eq:Ta}     \\
    {\cal T}^{b}_{\nu,\nu'}(\varepsilon)
&=&
    s_{\nu} (1-\delta_{\nu,\nu'})  {\left[a(\varepsilon)+a(\varepsilon-\hbar\omega)\right]}^2
+
    2s_{\nu}  \left( \delta_{\nu,\nu'}(1-2\gamma^2) + \gamma^2 \right)
    c(\varepsilon,\varepsilon-\hbar\omega)
    \left[ b(\varepsilon-\hbar\omega)+(1-2\delta_{\nu,\nu'}) b(\varepsilon) \right]  \nonumber  \\
&+& (2 \delta_{\nu,\nu'}-1) b(\varepsilon) b(\varepsilon-\hbar\omega)  \left[b(\varepsilon-\hbar\omega)-b(\varepsilon) \right]
{\left(\hbar \omega\right)}^2 \left(s_{\nu}\Gamma_{\nu}+s_{\nu'}\Gamma_{\nu'}\right)/\Gamma^3   \label{eq:Tb} \\
{\cal T}^{c}_{\nu,\nu'}(\varepsilon,\varepsilon')  &=&  \frac{4 }{\pi \Gamma}
\frac{s_{\nu} \Gamma_{\nu'} \hbar  \omega}{{\left(\hbar \omega\right)}^2-{\left(\varepsilon-\varepsilon' \right)}^2}
\left\{
a(\varepsilon) \left[ b(\varepsilon)+ b(\varepsilon') \right]  + 2 s_{\nu'} \gamma b(\varepsilon)
d(\varepsilon,\varepsilon')
\right\},
\label{eq:Tc}
\end{eqnarray}
\end{subequations}
\end{widetext}
are expressed through the functions
$a(\varepsilon)= \mbox{Re}[\Gamma G^{A}_{d,\,d}(\varepsilon)]$,
$b(\varepsilon)= \mbox{Im}[\Gamma  G^{A}_{d,\,d}(\varepsilon)]$,
$c(\varepsilon,\varepsilon')=
a(\varepsilon)a(\varepsilon')+b(\varepsilon)b(\varepsilon')$,
and
$d(\varepsilon,\varepsilon')=
a(\varepsilon)b(\varepsilon')-b(\varepsilon)a(\varepsilon')$,
while $s_{l/r}=\pm1$, $\gamma=(\Gamma_L-\Gamma_R)/\Gamma$ is an asymmetry factor for the coupling to the contacts, and
$n_B=[e^{\hbar\omega/k_B T} - 1]^{-1}$ is the Bose factor.
The two first terms in the r.h.s. of Eq.~(\ref{eq:I2}) express elastic as well as inelastic electron tunneling with emission/absorption of one phonon.
The last term cannot be interpreted as a single-particle elementary process:
It is related to the many-body character of the Fermi sea in the
leads~\cite{imam:1994,flensberg:2003,mitra:2004}.
In the following, we discuss the result for the current in several regimes.

Let us start with considering the linear conductance
$G \equiv {(\partial I/ \partial V)}_{V=0}=G^{(0)}+\alpha^2G^{(2)}+\dots$.
Far from the resonance, the tunnel regime is realized. Eqs.~(\ref{eq:I0})-(\ref{eq:transmission}) yield
$(h/e^2)G^{(0)}_\mathrm{tun}=4 \Gamma_l \Gamma_r/ \tilde\varepsilon_d ^2\ll 1$, where
$\tilde\varepsilon_d =\varepsilon_d - (\mu_l+\mu_r)/2$. From Eq.~(\ref{eq:I2}), we obtain
\begin{equation}
\label{eq:G2_T0_tun}
G^{(2)}_\mathrm{tun}=-4G^{(0)}_\mathrm{tun}
\left[  1  + 2 n_B -2\frac{\hbar \omega}{k_B T}n_B(n_B+1)\right] ,
\end{equation}
in agreement with Ref.~\cite{shekhter:2006}.
That is, the magneto-conductance is negative at $T=0$:
$\Delta G_\mathrm{tun}\equiv G_\mathrm{tun}-G^{(0)}_\mathrm{tun}=-4\alpha^2G^{(0)}_\mathrm{tun}$,
and it vanishes at high temperature like
$\Delta G_\mathrm{tun}/G^{(0)}_\mathrm{tun}=-4\alpha^2\hbar\omega/(3 k_B T).$

The general expression of $G^{(2)}$ at $T=0$ is
\begin{eqnarray}
G^{(2)}  \!\!& =& \!\! - G_\mathrm{max}
\!\!\!  \sum_{s=\pm}  \!\!\!
\left\{
 -\frac{s B_s(\mu)}{\pi}  \ln\left(
\frac{  \hbar^2 \omega^2}{\Gamma^2 + \tilde\varepsilon_d^2 }\right)
\right. \nonumber \\
& + &
\left.
A_{s}(\mu) \left[ 1  - \frac{2 s}{\pi} \arctan\left(
\! \frac{\tilde\varepsilon_d}{\Gamma} \! \right) \right]
+a^2(\mu) \right\}, \label{eq:G_T0}
\end{eqnarray}
where $A_\pm(\mu) =  a(\mu) a(\mu\pm\hbar\omega) + 2 \gamma^2 b(\mu) c(\mu,\mu\pm\hbar\omega)$
and $B_\pm(\mu)  = a(\mu) b(\mu\pm\hbar\omega) + 2 \gamma^2 b(\mu) d(\mu,\mu\pm\hbar\omega)$.
The correction $G^{(2)}$ is always negative.
It results in a narrowing of the resonance line at finite magnetic field.
The conductance peak at resonance ($\tilde\varepsilon_d=0$) is:
\begin{equation}
\label{eq:G2-peak}
G_{\mathrm res}^{(2)}
=
- G_\mathrm{max} \frac{4 \gamma^2 \Gamma^2}{\Gamma^2+{\left(\hbar \omega\right)}^2}
\left[ 1 + \frac{2 \hbar\omega}{\pi \Gamma} \ln \left( \frac{\hbar \omega}{\Gamma}\right)
\right].
\end{equation}
Remarkably the correction vanishes only for symmetric coupling between
the dot and the leads ($\gamma=0$),
in contrast with the polaronic problem where
$G^{(2)}_{\mathrm res}=0$ identically.
In the adiabatic limit, $\hbar\omega\ll\Gamma$, we find
\begin{equation}  \label {eq:G2T0_2}
G^{(2)} = - 4G_\mathrm{max}\Gamma^2[\tilde\varepsilon_d^2 +\gamma^2 \Gamma^2]/
[\tilde\varepsilon_d^2+\Gamma^2]^2.
\end{equation}
In the anti-adiabatic limit, $\hbar\omega\gg\Gamma$,
the dominant term contributing to Eq.~(\ref{eq:G_T0})  is
\begin{equation}
    \label{eq:G2-high-freq}
    G^{(2)}=-G_\mathrm{max}
    \Gamma^4/[\tilde\varepsilon_d^2+\Gamma^2]^2 \, .
\end{equation}
This result receives a simple interpretation~\cite{imam:1994}:
At high resonance frequency, the harmonic oscillator remains in its ground state, $|0\rangle$,
and an effective Hamiltonian for electron tunneling in the device is obtained by
projecting Eq.~(\ref{eq:Hamiltonian}) on this state.
This describes a noninteracting resonant level coupled to the leads
through renormalized tunneling matrix
elements $t_\nu\langle 0| e^{i\alpha_\nu x} |0\rangle=t_\nu e^{-\alpha^2/2}$
with corresponding level widths $\Gamma_\nu e^{-\alpha^2}$.
As a result, the resonance narrows, but the maximum transmission is unchanged.
Eq.~(\ref{eq:G2-high-freq}) expresses this in lowest order in the coupling constant.
The conductance reduction  arises to higher order in
$\Gamma/\hbar \omega$, $\Delta G_{\mathrm res}/G_\mathrm{max}\propto -\alpha^2\gamma^2\Gamma/(\hbar\omega)\ln(\hbar\omega/\Gamma)$.
%
%
%
%
\begin{figure}
\begin{center}
\includegraphics[scale=0.165,angle=270.]{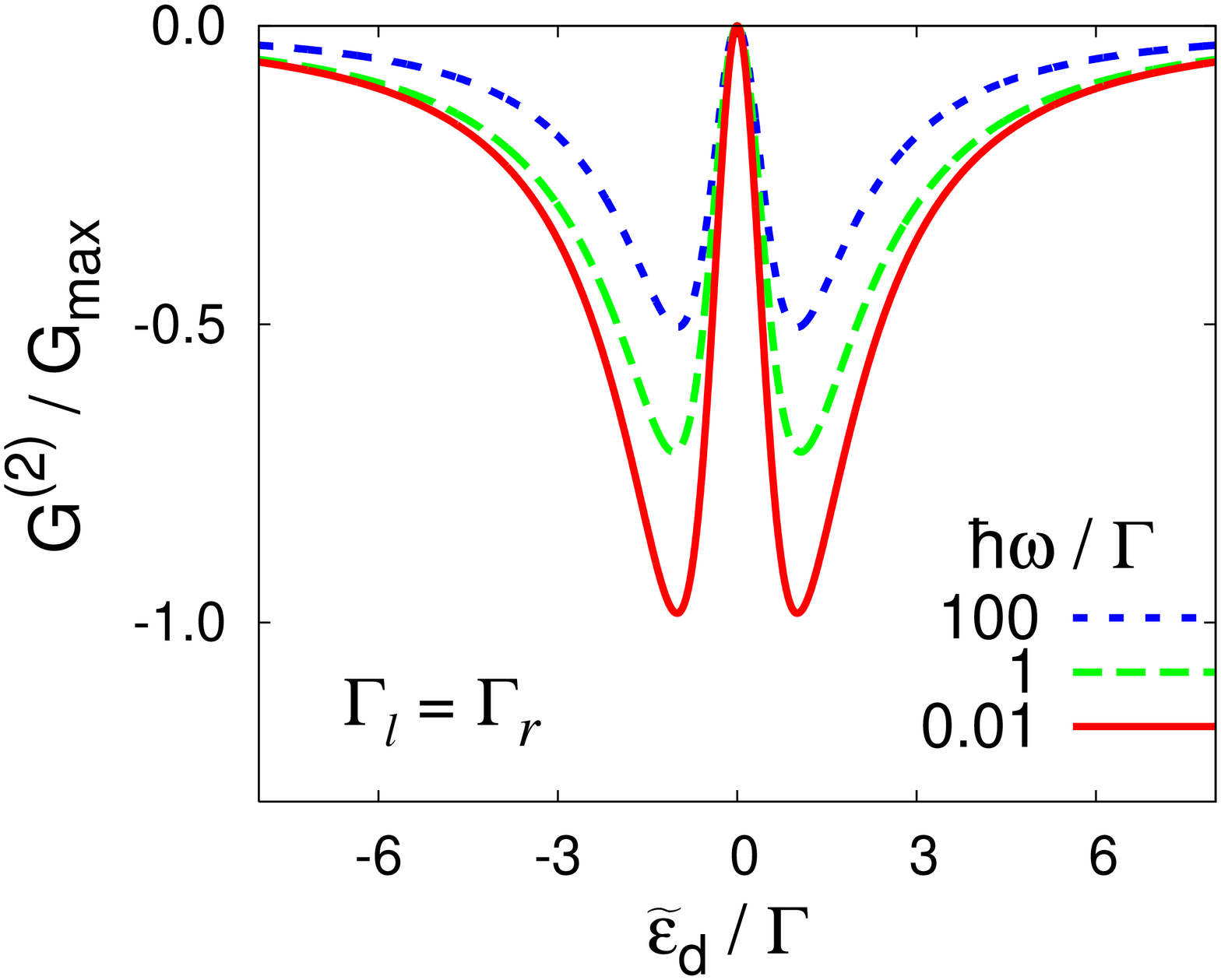}
\includegraphics[scale=0.165,angle=270.]{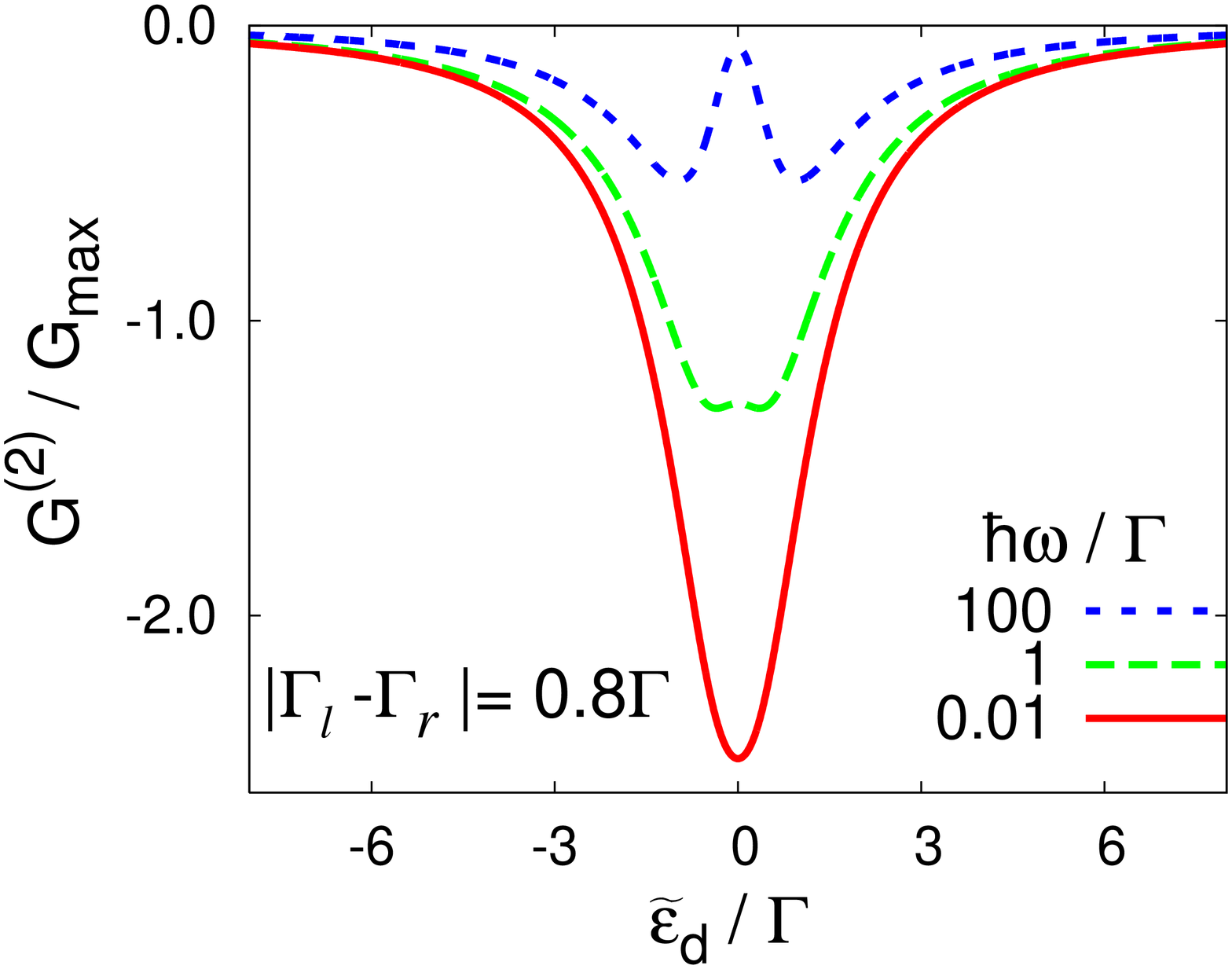}
\end{center}
\vspace{-6mm}
\caption{
Gate-voltage dependence of the magneto-conductance at $T=0$ and different values of
$\hbar\omega/\Gamma$, for a nanotube contacted symmetrically
(left) or asymmetrically (right).}
\label{fig2}
\end{figure}
The features discussed above are clearly visible in Fig.~\ref{fig2}.
As in the polaronic transport problem~\cite{mitra:2004}, there is no vibrational sideband in the gate-voltage dependence of the linear conductance.

It is worth mentioning that Eq.~(\ref{eq:G_T0}) coincides (up to order $\alpha^2$) with a Landauer-B\"uttiker formula
for the conductance in the elastic regime (zero bias-voltage and temperature) at arbitrary ratio $\hbar\omega/\Gamma$:
$G=(e^2/h)\sum_{kk'}|T_{kl,k'r}|^2\delta(\varepsilon_k)\delta(\varepsilon_{k'})$.
Here, the element of the T-matrix $T_{kl,k'r}=(g^{R}_{kl})^{-1}G^{R}_{kl,k'r}(g^{R}_{k'r})^{-1}$
is related to $G^{R}_{kl,k'r}$ that can be evaluated along the same lines as the Green's functions introduced above.
We believe that this result is not so trivial.
Indeed, contrarily to the polaronic Hamiltonian, the one that we are considering does not satisfy the ``proportionate coupling'' conditions to
 the right and left leads under which such a relation between conductance and transmission was proven in the presence of interactions~\cite{meir:1992}.

Does the effect discussed above really depend on the quantum vibrations of the oscillator?
>From Eq.~(\ref{eq:I2}), we find that the magnetoconductance is suppressed when $k_B T\gg\hbar\omega$:
\begin{equation*} \label{eq:GT}
\frac{\Delta G(T)}{G^{(0)}(T)}
=
-\alpha^2\frac{\hbar\omega}{k_BT}
\times
\left\{
\begin{array}{lr}
\frac{4}{3}
\frac{\tilde\varepsilon_d^2+\gamma^2\Gamma^2}
{\tilde\varepsilon_d^2+\Gamma^2},
&
k_BT\ll\Gamma,
\\
\frac{1}{2\cosh^{2}(\tilde\varepsilon_d/2k_BT)},
&
\Gamma\ll k_BT.
\end{array}
\right.
\end{equation*}
This result differs from the one obtained
with a classical description of the oscillator.
In this case, $x(t)$ obeys a classical equation
of motion in presence of a Langevin force
describing thermal fluctuations.
The displacement $x(t)$ acts as an external bias voltage
$V_\mathrm{ac}=2\alpha(\hbar/e)\dot{x}(t)$
in Eq.~(\ref{eq:HT}).
The current can then be calculated following Ref.~\cite{jauho:1994}.
This classical contribution coincides with
the term proportional to $n_B$ in the quantum result (\ref{eq:I2})-(\ref{eq:T}).
It can be viewed as a rectification of $V_\mathrm{ac}$-fluctuations.
We find that the corresponding magnetoconductance
$\Delta G_\mathrm{cl}$ is either negligible
(for $\hbar\omega \ll k_BT\ll \Gamma$) or different
(for $\hbar\omega,\Gamma\ll k_BT$ when it is of the same order)
from the quantum result.
In particular, in the tunnel regime,
$I(t)=G_\mathrm{tun}[V+2\alpha(\hbar/e)\dot{x}(t)]$
and $\Delta G_\mathrm{cl}=0$
since $\langle \dot x \rangle_\mathrm{cl}$ vanishes.

\begin{figure}[!hbtp]
\begin{center}
\includegraphics[scale=0.28,angle=270.]{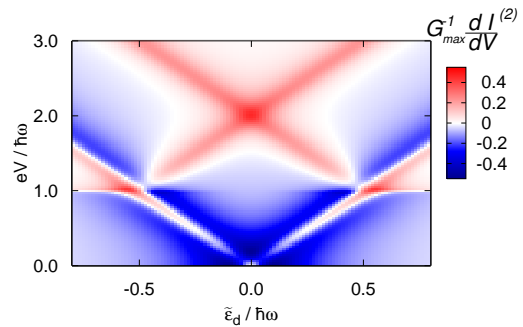}
\end{center}
\vspace{-6mm}
\caption{Gate and bias voltage dependence of the differential conductance
 $\partial I^{(2)}/\partial V$ at $T=0$ for $\hbar \omega=10 \Gamma$ and $\Gamma_r=\Gamma_l$.}
\label{fig3}
\end{figure}

The formula (\ref{eq:I2}) for the current also allows to address the finite bias nonlinear regime.
The possibility to excite the phonon (inelastic cotunneling) when $eV>\hbar\omega$ leads
 to a nonanalytical voltage dependence of the current in vicinity of the threshold at $T=0$, with
leading terms:
\begin{equation}
\frac{1}{G_\mathrm{max}}\frac{\partial I^{(2)}}{\partial V}=\frac{c_1}{2\pi}\ln\frac{\Gamma}{|eV-\hbar \omega|}
+4c_2\theta(eV-\hbar \omega),
\end{equation}
and asymptotic expressions for the coefficients
\begin{equation*}
c_1
=
\left\{
\begin{array}{lr}
-8 \hbar\omega
\Gamma^3
\left(\gamma^2 \Gamma^2 + \tilde \varepsilon_d^2\right)
/
\left(\Gamma^2 + \tilde \varepsilon_d^2 \right)^3,
&
\hbar\omega \ll \Gamma, \\
-\gamma (4 \Gamma/\hbar\omega)^4
\tilde\varepsilon_d/\Gamma,
&\hbar\omega \gg \Gamma,
\end{array}
\right.
\end{equation*}
and
\begin{equation*}
c_2
=
\left\{
\begin{array}{lr}
\Gamma^2 (\gamma^2 \Gamma^2 + \tilde\varepsilon_d^2)
/\left(\Gamma^2 + \tilde\varepsilon_d^2 \right)^2,
&
\hbar\omega \ll \Gamma, \\
\left( 2 \Gamma/\hbar\omega\right)^4
(\tilde\varepsilon_d^2 + \gamma^2\Gamma^2 )/\Gamma^2,
&\hbar\omega \gg \Gamma.
\end{array}
\right.
\end{equation*}
A similar feature has been discussed for the polaron problem~\cite{egger:2008}.
It is clearly seen in the gate and bias voltage dependence of the magneto-conductance shown in Fig.~\ref{fig3}.
In addition, Fig.~\ref{fig3} shows how the standard sequential tunneling lines
at $\tilde\varepsilon_d=\pm eV/2$ narrow at finite magnetic field,
while additional phonon-assisted tunneling lines appear at $\tilde\varepsilon_d=\pm (eV/2-\hbar\omega)$.

The weak-coupling approximation used above is well justified experimentally.
However, it overlooks several features that may deserve further study.
First, above the inelastic threshold ($eV>\hbar \omega$), the phonon mode is driven out of equilibrium.
By analogy with the polaron problem~\cite{mitra:2004,egger:2008}, we estimate that our results remain valid
beyond the threshold in a region of voltage bias $\delta V \ll \hbar \omega/e$ for $\hbar\omega \ll \Gamma$,
and $\delta V \ll  {\left( \hbar\omega \right)}^2/\Gamma e$ for $\hbar\omega \gg \Gamma$,
under the additional condition that the current-induced damping of the oscillator
$\alpha^2\min[(\hbar\omega)^3/\Gamma^2,\Gamma]$ remains much smaller than $ \omega$.
Second, additional inelastic lines, corresponding to multi-phonon absorption, appear at
$\tilde\varepsilon_d=\pm (eV/2-n\hbar\omega)$~\cite{boese:2001}
and lead to a magnetic-field-induced current suppression similar to the electrostatic
Frank-Condon blockade \cite{koch:2005} recently observed experimentally~\cite{leturcq:2009}.

In conclusion, we have studied how the Aharonov-Bohm phase accumulated by the electrons crossing
a vibrating nanotube affects its resonant magneto-conductance in the weak coupling regime.

We acknowledge important discussion with L. Glazman.
This work was supported by ANR through contract JCJC-036 NEMESIS.

\end{document}